\documentclass[12pt]{iopart}

\usepackage[utf8]{inputenc}
\usepackage{t1enc}
\usepackage[final]{graphicx}
\usepackage{times}
\usepackage{bm}

\begin{document}

\bibliographystyle{unsrt}

\title{Dipolar interactions and  thermal stability of two-dimensional 
      nanoparticle arrays}

\author{Daniel A Stariolo$^1$ 
\footnote{Research Associate of the Abdus Salam International 
Centre for Theoretical Physics, Trieste, Italy.}
and Orlando V Billoni$^2$ 
\footnote{Member of CONICET, Argentina}
}
\address{$^1$ Departamento de Física, Universidade Federal do Rio Grande do Sul\\
CP 15051, 91501-970, Porto Alegre, Brazil}
\address{$^2$ Instituto de Física de la Facultad de Matemática, Astronomía y Física 
(IFFAMAF-CONICET),
Universidad Nacional de Córdoba\\ 
Ciudad Universitaria, 5000 Córdoba, Argentina}
\ead{stariolo@if.ufrgs.br}
%

\begin{abstract}
We show results of Monte Carlo simulations of an array of monodispersed magnetic monodomain
particles, in a square lattice with dipolar interactions and perpendicular uniaxial 
anisotropy. We first show the equilibrium phase diagram of the system, which shows three phases,
superparamagnetic, out-of-plane antiferromagnetic and in-plane antiferromagnetic
with a reorientation transition between the last two. We then employ a recently
introduced Time Quantified Monte Carlo method to study the relaxation of autocorrelations 
of the particles array for different ratios of dipolar to anisotropy energies. 
In the non-interacting case we show that relaxation is exponential in
time with characteristic times obeying a classic result by Brown. 
When dipolar interactions are switched on, the relaxation is very well described
by stretched exponential forms in the whole time window and in both the
superparamagnetic and ordered phases. Relaxation times still obey a nearly Arrhenius 
behavior, with a single effective energy barrier that decreases as the dipolar interaction 
increases, a result that must be interpreted within the dynamics protocol. 
No signs of glassy behavior were found, in agreement with the absence
of disorder in the model system. 

\end{abstract}



\submitto{\JPD}
\maketitle

\section{Introduction}

The development of advanced experimental tools to fabricate magnetic nano-sctructures in a
controlled way
has led to an enormous growth of studies concerning the physical understanding of nanomagnetism, 
such as the role of different interactions and scaling behavior. Nanostructured magnetic 
materials like magnetic particles and patterned magnetic alloys are  very important 
systems in the experimental and theoretical research \cite{Jo2004}, beyond of  its  important 
technological applications like magnetic data  storage and others fields \cite{Ro2001,ReHu2005}. 
In general, magnetic media for data storage  are composed of tiny but isolated magnetic 
nanocrystalline grains, but the need of increasing the
memory densities forces the development of non-conventional media, like patterned media and 
self-assembled magnetic nanoparticles.  
The idea is to replace the  randomly oriented magnetic grains with magnetic particles or nanodots 
in which it is possible to control
the magnetic anisotropy orientation \cite{ReHu2005, Ro2001} with the advantage of reducing the
 noise. Whatever the magnetic support,  thermal stability is a key ingredient in  magnetic 
media for data storage. 
Increasing the storage density implies a reduction in the size  of the magnetic grains.
If the particles or grains are too small they  lose thermal stability reaching the
so called {\it superparamagnetic limit}. For single isolated  magnetic nanoparticles,  
modelled as giant magnetic moments, the thermal 
stability problem has been  extensively studied from the theoretical point of view 
\cite{Br1963,Garcia-Palacios98PRB,Coffey98PRL}, however, when the system is 
a set of interacting  particles there are still many open questions 
\cite{MaSa2006, SaHyMeJaMeGiChMaFe2001, DeTr2001}. 
In particular, long-range dipolar interactions are unavoidable becoming  relevant when 
increasing the packing density of the magnetic moments. In this sense, it is very important 
the understanding of the role of dipolar interactions in the relaxation dynamics
of a set of magnetic moments due to thermal fluctuations.  
Patterned alloys \cite{Ro2001} and two dimensional self-assembled  magnetic nanoparticle
arrays are systems suitable for the experimental 
study of thermal stability, since it is possible to tailor the magnetic anisotropy orientation 
of the magnetic nanoparticles with respect to the array plane. In addition, other important 
properties like the shape and size of the magnetic units can also be controlled.
A landmark of magnetic relaxation in particle arrays is the slow, quasi-logarithmic decay
of the remanent magnetization over several decades in time. The origin of this slow
relaxation is still controversial. Recent simulation studies in three
dimensional systems point to the necessity of considering very
simplified models, with minimum ingredients, as a way to understand
the mechanisms responsible for the slow relaxation and glassiness in
dipolarly coupled magnetic
particles~\cite{GaPoRiBu2000,UlGaRiBu2003,Po2005,RuBu2006,RuBu2007}.
The structure of the crystal lattice (or the absence of structure),
the nature of anisotropies (whether random or not), the presence of
polydispersity in the particle volumes, and the volume concentration of
the array, all have to be carefully considered in order to better
understand the origin of magnetic behavior in a particular array.

The relaxation behavior of two dimensional arrays of monodispersed magnetic moments 
has been studied theoretically 
\cite{LoWhDa1991,SuLu1997,SaHyMeJaMeGiChMaFe2001,DeTr2001,DeLyTr2003,DeLyTr2004}. 
Mean field calculations \cite{LoWhDa1991} show that the relaxation may be slow without the 
need of introducing disorder, as commonly assumed.
The mean field approach does not take into account correlations between particles due to 
dipolar interactions and rapidly fluctuating 
thermal fields  \cite{DeTr2001}. In fact, when dipolar interactions are considered, there are 
ordered states at low temperatures. Taking as example an array of
particles in a square lattice and depending on the ratio between the dipolar 
and anisotropy  strength, the system can order antiferromagnetically in the plane or out of 
the plane \cite{MaWhDeHo1996,CaBiPiCaStTa2008}. 
The complexity of the interacting problem requires the use of numerical tools in order to go
beyond mean field results \cite{DeLyTr2003}. 

In this work we first show, by means of Monte Carlo simulations,  the phase diagram of a system 
of monodispersed magnetic particles in
a square lattice with perpendicular uniaxial anisotropy and dipolar interactions. With this
information at hand, we then use a recently introduced Time Quantified Monte Carlo method to 
characterize
 the relaxation dynamics of the system. We characterize the time
 dependent relaxation near thermal
equilibrium, which
happens to be very well described by stretched exponential decays in the whole time
window and in both ordered and disordered regions of the phase diagram. From the relaxation curves we go on to obtain the
characteristic relaxation times and compare them with the predictions of the Brown-Arrhenius
theory. In order to explore the influence of dipolar 
interactions on the effective free energy barriers that control the relaxation process,
we explore a wide range of intensities of the dipolar interaction. 
This allows us to study the limits of validity of the Brown-Arrhenius model of relaxation and
define the basic ingredients responsible for the observed slow dynamics.

\section{Model and Simulations}

We consider a model of a two dimensional monodispersed array of single domain magnetic nanoparticles arranged in a square lattice. 
The particles have uniaxial anisotropy  which is oriented perpendicular to the $x-y$ plane of the array. The classical Hamiltonian 
of this system can be written in the form: 
\begin{equation}
H =  - D \sum_{i}S_{iz}^2 + g \sum_{i<j} \frac{\vec{S}_i \cdot \vec{S}_j -
3(\vec{S}_i \cdot \hat{e}_{ij}) (\hat{e}_{ij} \cdot \vec{S}_j) }{r_{ij}^3},
\end{equation}
\noindent where $\vec{S}_i=\vec{\mu}^{\,i}_{np}/\mu_{np}$ are three-dimensional magnetic moments 
of unit length. $\mu_{np}$ is the value of the magnetic moment, which depends of the 
volume $V$ and the magnetization $M_s$ of the magnetic nanoparticle, $\mu_{np}=V M_s$. 
$\hat{e}_{ij}=\vec{r}_{ij}/r_{ij}$ are unit vectors on the plane of the array.
The first term is the energy  contribution of the uniaxial anisotropy pointing in the direction of the $z$ axis. 
$D=K_u V$ with $K_u$ the uniaxial anisotropy constant of the system under study. 
The second term in the Hamiltonian is the dipolar  interaction between the magnetic particles, 
$g = \frac{\mu_0 \mu_{np}^2}{4\pi a^3}$ where $\mu_0$ is the vacuum permeability  and $a$ is the lattice parameter. 
We consider the dipolar energy in a point dipole approximation~\cite{PoPi2002}.  
In order to obtain the equilibrium phase diagram of the system we employ the 
Metropolis Monte Carlo algorithm where the spins are randomly updated in the unit 
sphere~\cite{Nowak01ARCP}. For the simulations of the relaxation dynamics we 
implemented a recently introduced Time Quantified Monte Carlo Method (TQMC)
~\cite{Nowak00PRL,ChJaLeOk2006PRL,ChJaLe2006PRB} 
which allows direct comparison with experimental time scales. We considered the high 
damping limit in order to maximize the total time span of the simulation. 
The simulations were done for square lattices of linear dimensions
$L=16$ and $L=32$. 
In order to minimize
finite size effects due to the long ranged dipolar interactions, periodic 
boundary conditions were imposed by means of the Ewald method. Details 
for the implementation of the Ewald method can be found in ref. \cite{We2003}.
Once the dipolar and anisotropy effective fields are calculated 
for each magnetic moment in the lattice, they are updated randomly using 
the TQMC \cite{ChJaLe2006PRB}.

\section{Results}

\begin{figure}
\begin{center}
\includegraphics[height=7cm,width=10cm]{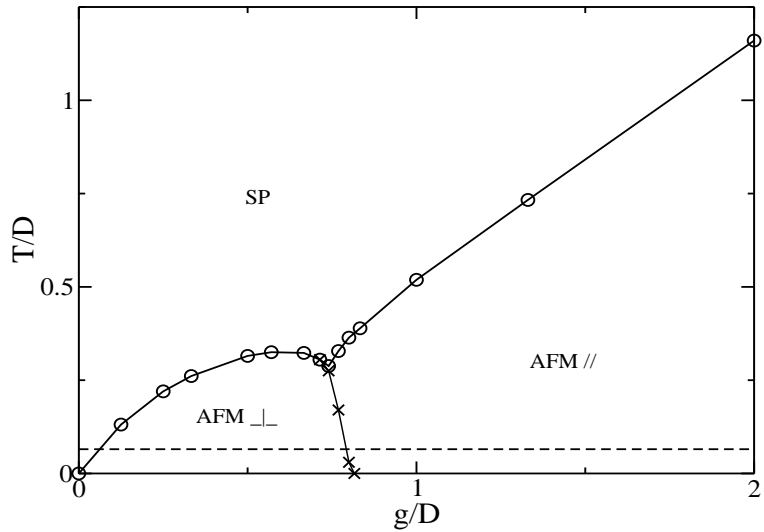}
\caption{\label{phasediagram} Phase diagram obtained from Monte Carlo Simulations for a 
system size L=32. The dashed line indicates the blocking temperature for the case of 
non-interacting particles.}
\end{center}
\end{figure}
\subsection{Phase diagram}
Figure \ref{phasediagram} shows the phase diagram of the system which was 
obtained and discussed in the context of the spin reorientation transition 
in reference~\cite{CaBiPiCaStTa2008}.
Here we plot it in a way suitable for discussing the dynamical behavior. All
energy scales are in units of the Boltzmann constant, which was set to $k_B=1$.
The dashed line in the figure corresponds to the blocking temperature for our 
simulated time window, for the case of an array of non-interacting particles, 
i.e $g/D=0$. 
For high temperatures the particles are superparamagnetic (SP phase). 
Upon lowering $T$ the anisotropy and dipolar terms in the energy begin to rule the 
behavior of the system, leading to two possible phases with antiferromagnetic order: for small temperatures 
and $g/D < 0.8$ the film orders out of plane in a checkerboard like configuration 
\cite{AiMeLeJaFeMaChGiViRoLaBe1998}. The perpendicular staggered
magnetization shows a
jump at the transition temperature, suggesting that the transition to
the
perpendicular antiferromagnetic phase is
first order. Of course, a precise determination of the order of the
transition would only be possible considering much larger system
sizes, which is beyond the scope of this work.
Our results in the present study all refer to this region
of the phase diagram.
For comparison, from an array of 4 nm Co particles we obtain $g/D=0.67$  or $g/D=0.084$, 
if the particles  are in contact or are separated by a distance equal to the diameter, respectively. 
Here we use the Co parameters of Ref. \cite{BaAlWePaWeGaSi99S}.  With growing dipolar 
coupling, the system goes through a spin reorientation transition to
an in-plane antiferromagnet as described in \cite{CaBiPiCaStTa2008}.

The knowledge of the different phases in an interacting system is essential to 
correctly interpret the relaxation behavior which will be the subject of the
following sections.
 
\subsection{Near equilibrium relaxation}

For the dynamical simulations we employed a recently introduced Time
Quantified Monte Carlo Method (TQMC)~\cite{Nowak00PRL,ChJaLeOk2006PRL,ChJaLe2006PRB} which allows to map Monte 
Carlo time steps to real time scales for any values of the damping parameter
usually introduced in micromagnetic simulations. This new TQMC is very
efficient and much more easy to implement and control than a
micromagnetic simulation. The new approach is based in a controlled
map between the stochastic Monte Carlo dynamics and the corresponding
stochastic Fokker-Planck equation. By means of a detailed
correspondence between the two approaches to the same stochastic
dynamics, the real time scale (in units of a characteristic {\em damping time})
and MC time steps become related by:
\begin{equation}
\Delta t[\tau_K] = \alpha \, \frac{R^2}{20} \, \Delta t[MCS],
\end{equation}
where the damping time is given by:
\begin{equation}
\tau_K = \frac{1}{\gamma H_k} \frac{(1+a^2)}{a}.
\end{equation}
In the previous relations $\alpha=D/T$, $R$ is the size of a cone where the spin is updated, 
$H_k=2K/M_s$, $M_s$ is the saturation magnetization and $a$ is the damping parameter. 
For details on the method and notations see references~
\cite{ChJaLeOk2006PRL,ChJaLe2006PRB,BiSt2007}. In the present
simulations $0.03 < R < 0.1$ in order to optimize the time span of the
dynamics for the different temperatures.

In order to characterize the relaxation dynamics of the system as
  it approaches equilibrium,
we measured time dependent auto-correlation functions by first relaxing the system  for 
$10^5$ MC steps 
from an initially disordered configuration of the spins. Then an initial 
configuration 
was stored \{${\vec S}_i(0)$\} and subsequent auto-correlations were defined as:
\begin{equation}
C(t) = \frac{1}{N} \sum_{i=1}^N {\vec S}_i(0) \cdot {\vec S}_i(t).
\label{correlation}
\end{equation}
Statistical averages of this function for many realizations of initial
conditions were done.
Auto-correlations relax similarly to the magnetization and are related to the linear
response function by the fluctuation-dissipation theorem. 

\begin{figure}
\begin{center}
\includegraphics[width=10cm,angle=0]{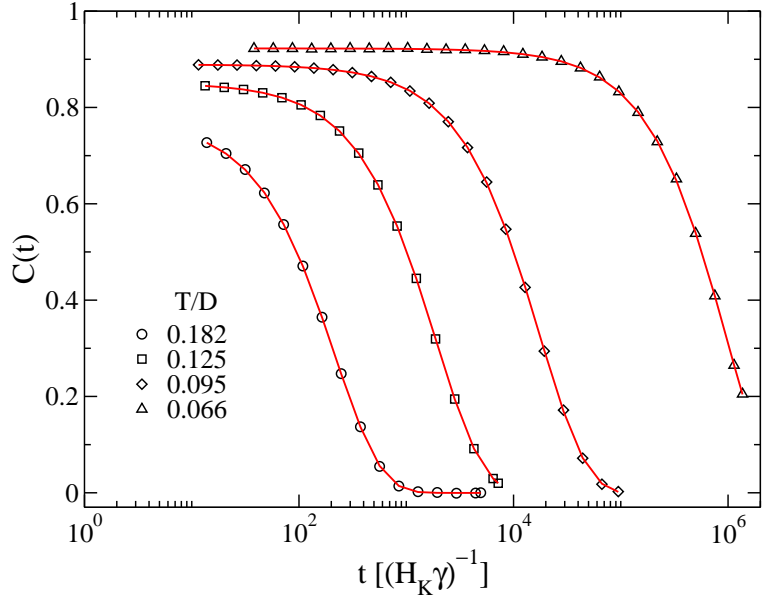}
\caption{\label{cg0} Autocorrelations for a system of noninteracting
particles. Continuous lines are exponential fits.}
\end{center}
\end{figure}

Our reference system is an ensemble of noninteracting particles
with uniaxial anisotropy. This corresponds to the vertical axis of
figure (\ref{phasediagram}).
In figure (\ref{cg0}) we show four correlation curves for this case,
each curve corresponding to a low temperature in the high energy
barrier regime, $D/T \gg 1$. The correlation curves show a rapid decay
to a rather high value where a plateau is developed as the temperature
is lowered. This regime corresponds to the relaxation inside the
initial basin of the energy minimum near the initial condition. At
long times the particles are able to overcome the anisotropy barriers
and a final relaxation takes place where the particles
decorrelate from their initial state. In this case of noninteracting 
particles the relaxations are exponential in time, 
as can be seen in the fits in
figure (\ref{cg0}):
\begin{equation}
C(t) = C_0 \, e^{-\frac{t}{\tau}}.
\end{equation}

\noindent Note that the horizontal time scale is given in units of $(\gamma \,H_k)^{-1}$. 
Taking as reference cobalt with a  high damping constant $a=1$, then $\tau_K=4.82 \times 10^{-11}\,s$ 
using $\gamma = 0.2212 \,10^{6}\, m/A s^{-1}$ for the gyromagnetic ratio   
and $H_k=168\, kA/m$ for the anisotropy field \cite{BaAlWePaWeGaSi99S}.

From the results for the correlation functions we can obtain the
relaxation times for the particles. In a classic paper, Brown obtained
approximate expressions for the relaxation time of single domain magnetic
particles. In the case of zero applied field and high energy barriers
the result of Brown is~\cite{Br1963}:
\begin{equation}
\frac{\tau}{\tau_K} \approx \frac{\sqrt{\pi}}{2} \, \alpha^{-1/2} \,
e^{\alpha}.
\label{tau_brown}
\end{equation}

\begin{figure}
\begin{center}
\includegraphics[height=12cm,width=8cm,angle=-90]{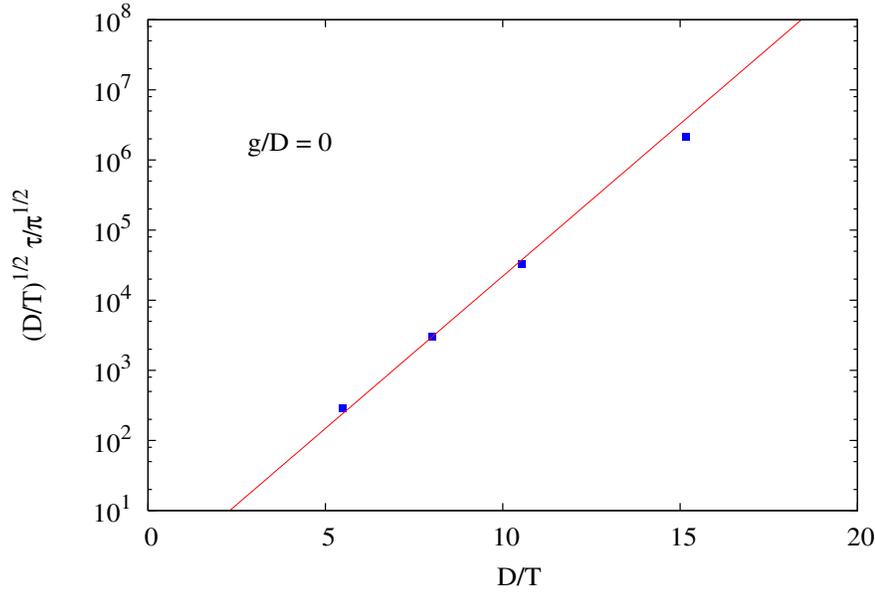}
\caption{\label{tg0} Relaxation times for a system of noninteracting
  particles. The solid line is the analytic approximation by Brown.}
\end{center}
\end{figure}

In figure (\ref{tg0}) we show the relaxation times obtained from the
exponential fits of figure (\ref{cg0}) together with the analytic
approximation by Brown in a log-linear plot. The agreement is very good except for the
point corresponding to the lowest temperature. This departure may be
due to insufficient  relaxation of the particles for this very high
energy barrier. This explanation can be supported by
calculating the {\em blocking temperature} for the ensemble of
particles, defined as the temperature for which the experimental time
scale equals the relaxation time. This $T_B$ is usually
calculated assuming an exponential or Arrhenius behavior of the
relaxation time. This is not completely true in this case, where the
prefactor of the exponential in (\ref{tau_brown}) depends on
temperature. Nevertheless a fit with a purely Arrhenius behaviour of the form
$\tau = \tau_0 \exp{(\Delta E/T)}$ works reasonably well and allows a rough
determination of the blocking temperature:
\begin{equation}
T_B = \frac{D \Delta E }{\ln{(t_{exp}/\tau_0)}},
\end{equation}
where in our simulations $t_{exp} \simeq 10^6$. From the fitting values
of $\Delta E=0.85$  and $\tau_0=2.27$, we obtain $T_B/D \approx
0.065$ which is near the value of the lowest temperature
simulated. Then it is probable that the curve corresponding to $T=0.066$ in
Fig. (\ref{cg0}) be still  subject to considerable fluctuations.

\begin{figure}
\begin{center}
\includegraphics[width=10cm,angle=0]{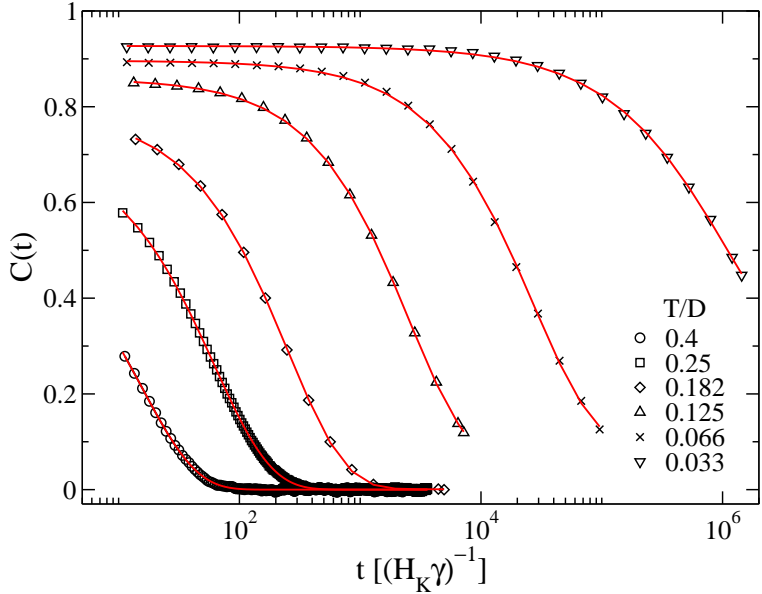}
\caption{\label{cg00625} Correlations for $g/D=0.0625$. Solid lines are
fits to stretched exponential forms (see text).}
\end{center}
\end{figure}

Upon switching the dipolar interaction the main observation is that the relaxation
is no more exponential. In figure \ref{cg00625} we show correlation curves
for a small intensity of the dipolar interactions $g/D=0.0625$ and different
temperatures, all in the superparamagnetic phase (see figure \ref{phasediagram}).
Together with the data points, the solid lines represent fits with a {\em
stretched exponential form}:
\begin{equation}
C(t) = C_0 \, e^{-(\frac{t}{\tau})^{\beta}}.
\end{equation}
In this case of weak dipolar interaction the values of the exponent $\beta$ are
around $0.9$. In fact we found that the typical $\beta$ values diminish as
the intensity of the dipolar coupling grows. 
Moreover, the stretched exponential form can describe the whole time
span of the relaxation for any value of the dipolar intensity. Nevertheless,
to our knowledge, there is no analytic prediction for this particular
form of the time dependent relaxation in systems of interacting particles.

A quasi-logarithmic time decay is widely employed to fit relaxation data
in systems with some kind of disorder. Nevertheless some time ago Dahlberg 
et al.~\cite{DaLoWhMaEn1994} showed that this slow relaxation need not be
associated with disorder in the system and can as well be present due to
the long ranged dipolar interactions. They verified this by solving 
numerically a mean field model of two state particles. More recently 
Denisov et al.~\cite{DeTr2001} obtained analytic expressions for the relaxation
of the remanent magnetization in a 2d system of interacting dipoles with
strong perpendicular anisotropy from an initially saturated
state. They
find a crossover from slow relaxation at intermediate times to
exponential one at very long times. Although the models are similar, a
direct comparison of their results with ours is not straightforward
due to the different protocols used in both cases.

\begin{figure}
\begin{center}
\includegraphics[width=10cm,angle=0]{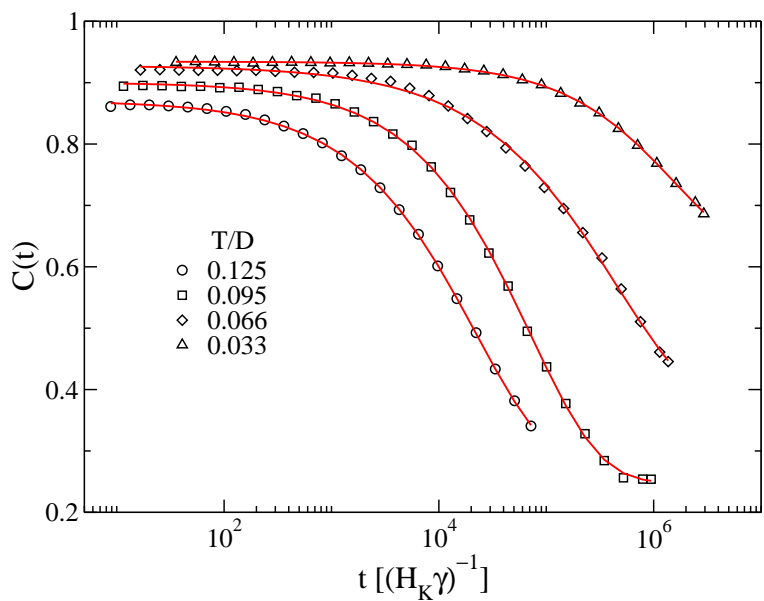}
\caption{\label{cg025} Correlations for $g/D=0.25$. Solid lines are
fits to stretched exponential forms (see text).}
\end{center}
\end{figure}

The stretched exponential relaxation is observed also in the low temperature
regime of the present system, when it relaxes towards a state
with long range antiferromagnetic order out of the
plane. In figure (\ref{cg025}) we show autocorrelation curves for temperatures
corresponding to the ordered phase. 
For this value of $g/D=0.25$ the stretching exponent $\beta$ is around $0.6$,
i.e. the relaxation is much slower than in the non-interacting case.
Nevertheless, although not shown in the figure, the behaviour of relaxation
times does not present any anomaly when crossing the phase transition
temperature. Due to the first order nature of the transition (and the
small size of the system) only a gradual change of stability between the
disordered and ordered phases takes place, with no evident effect on
the typical relaxation times. 

\begin{figure}
\begin{center}
\includegraphics[width=8cm,angle=-90]{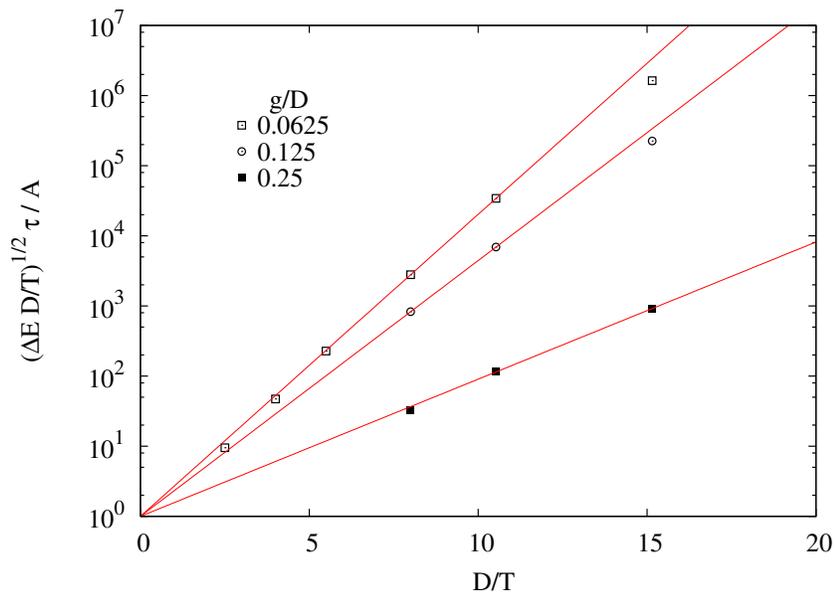}
\caption{\label{todos} Relaxation times for a system with three different 
dipolar to anisotropy interaction ratios. The solid lines are fits with 
expression (\ref{taus_fits}) as described in the text.}
\end{center}
\end{figure}

Together with the departure from exponential decay, when interactions are switched
on the relaxation times also behave differently than the independent particle
relation eq. (\ref{tau_brown}). In figure (\ref{todos}) the characteristic relaxation
times corresponding to dipolar to anisotropy ratios $g/D=0.0625, 0.125$ and $0.25$
are shown. The
values of $\tau$ were obtained from the corresponding stretched exponential
fits and the data was fitted with the functional form:
\begin{equation}
\frac{\tau}{\tau_K} \approx \frac{A}{2} \, (\alpha \,\Delta E)^{-1/2} \, e^{\alpha \,\Delta E},
\label{taus_fits}
\end{equation}
\noindent which is a natural modification of Brown's relation (\ref{tau_brown}),
where now  $D \Delta E$  represents an effective free energy barrier (we fixed $D=1$ in the
simulations whenever $g \neq 0$). Typical values of
$\Delta E$ are $0.99$, $0.83$ and $0.45$ for $g/D=0.0625, 0.125$ and $0.25$ respectively.
It was not possible to fit the data for the interacting systems fixing $\Delta E$ to
one, as in the noninteracting case, even considering the exponent of the temperature
dependent prefactor as a fitting parameter different from 1/2.
It is important to note that while the data for $g/D=0.0625$ all correspond to the
superparamagnetic phase, those corresponding to the two larger values of dipolar interaction
probe the system relaxing towards an ordered phase. In spite of this, it can be seen in figure
(\ref{todos}) that in all cases the data is compatible with a single free energy
barrier as represented by the nearly Arrhenius form (\ref{taus_fits}). From this
behavior it can be concluded that glassy behaviour is not present in the dipolar
interacting system, as this case should imply a non-Arrhenius dependence of relaxation
times.  Of course, some kind of disorder not considered in the present
simulations, e.g. random anisotropies or polydispersity,
may lead to spin glass like behaviour (see, for example, discussions in
\cite{UlGaRiBu2003,Po2005} and references therein).
Instead, we observe that dipolar interactions  are responsible
  for an  effective reduction of the  free energy 
barriers which control the relaxation process, even with an initial state where
demagnetization fields are 
significantly reduced. 
This energy barrier reduction is related to the growing of the ordered
state  as can be realized by 
looking at the order parameter evolution during the relaxation. 
It is worth to stress that the growing of the ordered phase does not
affect the Arrhenius character of  the relaxation process.

\section{Conclusions}

We have studied the relaxation dynamics of an array of dipolar interacting
magnetic particles. Instead of looking at the remanent magnetization
from a saturated state, we have first let the system relax near
equilibrium and the subsequent relaxation was studied. Our model of a
monodispersed array of single domain particles with uniform
perpendicular anisotropy, while clearly simplified, is a first step
towards a more systematic study of the origin of slow dynamics in
realistic nanoparticles arrays. We obtained two main results: first,
regarding the time dependence of relaxation functions, we found that
while the relaxation is exponential in the noninteracting case, it
changes to a slower relaxation form as the dipolar interaction is
switched on, already for a very weak interaction strength. A stretched
exponential form fits remarkably well the data over the whole time
span of the relaxation, contrary to the known limitations of a
logarithmic fit~\cite{LoWhDa1991,Ah2000,IgLa2004}. 
While this may indicate a fundamental role of the
stretched exponential form as a consequence of the microscopic
interactions
~\cite{GaPoRiBu2000,UlGaRiBu2003,Po2005}, this remains to be proved on theoretical basis. We also found
that the slow dynamics is present in the interacting system regardless
of the temperature regime, while  relaxing towards the superparamagnetic or ordered
phases. 
Second, we
showed that the Brown-Arrhenius result for the temperature dependence
of relaxation times works very well for independent particles, but
needs to be modified when interactions are present. We proposed a
slight modification of the original Brown's results in order to fit
the data. Within this form, the behaviour of relaxation times shows
that effective free energy barriers are reduced  with growing
dipolar interactions, at least in the present case where the system is initially
already in a demagnetized state.

The Time Quantified Monte Carlo method turned out to be a convenient
and efficient way to perform simulations which can be directly
confronted with experimental data. We used the example of cobalt
nanoparticles in which we showed that realistic time scales can be
reached within reasonable model parameters. 

 Monte Carlo simulations of
three dimensional arrays of particles, including both configurational and 
anisotropy disorder
\cite{GaPoRiBu2000,UlGaRiBu2003,Po2005,RuBu2006,RuBu2007}, have shown 
that anisotropy disorder smears out the effect of configurational
disorder \cite{Po2005},
 and that this disorder is particularly
important when the ground state is a columnar antiferromagnet
\cite{Po2005,RuBu2006}.
Moreover, glassy behavior, present in these cases,
seems to be enhanced in the configurationally ordered case
\cite{RuBu2007}.
Then, future
studies of two dimensional systems should add the effects of dipolar interactions
with the influence of disorder on relaxation, present for example 
when the anisotropy axes are randomly oriented, which is the usual
case in self-assembled or epitaxial nanoparticle systems
~\cite{IgLa2004,FiSc2008,ZhWeXiSu2003}. 
For the important case of perpendicular anisotropy, relevant for data storage, disorder
can also be present in the intensities of local
anisotropies~\cite{Ly2000,ZhWeXiSu2003}. 
Polydispersed assemblies with a distribution of grain
volumes can also be straightforwardly considered within the present
model~\cite{IgLa2004,FiSc2008}. We plan to address this issues in future works, 
together with
different dynamical protocols more directly related with experiments,
as for example, the behaviour of the thermoremanent magnetization in field cooled 
protocols and the influence of
external fields after a zero field cooled process.

\ack
This work was partly supported by CNPq (Brazil) and a binational program CAPES/SeCyT through 
grant $N^o 098/06$.
Partial support was also provided by grants from CONICET, SECyT-UNC and FONCyT grant PICT-2005 
33305 (Argentina).


\end{document}